\documentclass[]{emulateapj_mod}
\usepackage[active]{srcltx}
\usepackage{graphicx}
\usepackage{apjfonts}
\newcommand{\myemail}{manojendu@ncra.tifr.res.in}
\newcommand{\grs}{GRS~1915+105}
\newcommand{\x}{Cygnus~X-3}
\shorttitle{Hard X-ray time lag: GRS 1915+105}
\shortauthors{Choudhury et al.}
\begin{document}

\title{Anti-correlated hard X-ray time lag in \grs: evidence for a truncated accretion disc}
\author{Manojendu Choudhury\altaffilmark{1}}
\affil{National Centre for Radio Astrophysics, T.I.F.R., Pune-411007. India}
\author{A. R.Rao, Surajit Dasgupta}
\affil{Tata Institute of Fundamental Research, Mumbai-400005. India}
\author{J. Pendharkar, K. Sriram} 
\affil{Department of Astronomy, Osmania University, Hyderabad, India}
\author{V. K. Agrawal}
\affil{Inter University Centre for Astronomy and Astrophysics, Pune-411007. India}
\altaffiltext{1}{e-mail: \myemail}

\begin{abstract} Multi-wavelength observations of Galactic black hole candidate sources indicate a close connection between the accretion disk emission and the jet emission. The recent discovery of an anti-correlated time lag between the soft and hard X-rays in \x~\citep{choudhury04apjl} constrains the geometric picture of the disk-jet connection into a truncated accretion disk, the truncation radius being quite close to the black hole. Here we report the detection of similar anti-correlated time lag in the superluminal jet source \grs. We show the existence of the pivoting in the X-ray spectrum during the delayed anti-correlation and we also find that the QPO parameters change along with the spectral pivoting. We explore theoretical models to understand this phenomenon.

\end{abstract}

\keywords{accretion -- binaries : close -- stars : individual (\grs) -- X-rays : binaries}

\section{Introduction}
In the current era, discerned by the wide band X-ray spectral capability of the
present day X-ray satellite observatories, the non-thermal component in the hard
X-ray spectrum has emerged to be the most prominent and intriguing observational
feature of the Galactic black hole candidates 
\citep[see][for a short review]{barret04}. Various conflicting hypotheses concerning
the physical mechanism of the origin of this non-thermal emission have been offered
to explain the phenomenon; ranging from synchrotron at the base of the jet
\citep{markoff01aa, markoff03aa} to Comptonization of thermal photons by a hot corona
with various geometrical structures \citep[see, for eg.,][]{poutanen98} -- the most
favored being that of a hot quasi-spherical cloud inside a truncated disc
\citep{zdziarski02apj}. In recent times, the existence of a truncated accretion disk
near black holes is establishing itself as an emerging independent paradigm, with
diverse theoretical formalisms, viz. ADAF \citep{narayan94apjl} and TCAF
\citep{chakrabarti96phr}, requiring the given geometrical structure.
The detection of an anti-correlated delay ($\lesssim$ 1000 s) of the hard X-ray
(20--50 keV) emission with respect to the soft X-ray (2--7 keV) emission in the X-ray
hard state of the enigmatic X-ray binary \x~\citep{choudhury04apjl}, giving rise to
the pivoting behaviour in the spectrum \citep{choudhury02aa}, adds credence to this
paradigm of truncated accretion disc, with the inner region consisting of
(quasi-)spherical flow of highly energetic matter Comptonizing the soft seed thermal
photons from the outer disc. The time scale of this delay may be attributed to the
viscous time scale of flow of matter in the radiation pressure dominated optically
thick accretion disk.

\grs~\citep{castro92iauc} is a   black hole binary system which displays the most
varied types of emission states ever seen in a Galactic microquasar, with the time
scale of spectral variations/transitions ranging from minutes to months.
Among all the various classes of behavioural features exhibited by \grs, in the
$\chi$ state \citep{belloni00aa} it exhibits steady X-ray emission for long
durations, further characterized by  a pronounced QPO feature. \citet{mcclintock04}
have examined the spectral classifications of black hole sources and have concluded
that \grs, in its steady states, shows characteristics closer to the Thermally
Dominant state or Low Hard sates. The $\chi$ state, with its band limited power
density spectrum, resembles the characteristics of the Low Hard state (though the
luminosity and the X-ray spectral index are slightly different from those found in
other black hole sources). In this state the X-ray spectra display a pivoting
behaviour (for spectra obtained on different days, spanning the extent of the hard
state), very similar to \x~\citep{choudhury03apj}. Also, the soft X-ray flux is
correlated to the radio emission in this state, with the correlation spanning across
5 orders of magnitude of intrinsic luminosity for various black hole candidates
\citep{choudhury03apj}, suggesting that the soft X-ray flux is driving both the hard
X-ray as well as radio emission. Given the universality of the correlated behaviour
of these systems, and the similarity of the spectral evolution in this state of
\grs~with \x, it was imperative that an anti-correlated delay between the hard and
soft X-rays, analogous to \x, be present in \grs, in the $\chi$ state.

\begin{figure}
\figurenum{1}
\epsscale{1.00}
\plotone{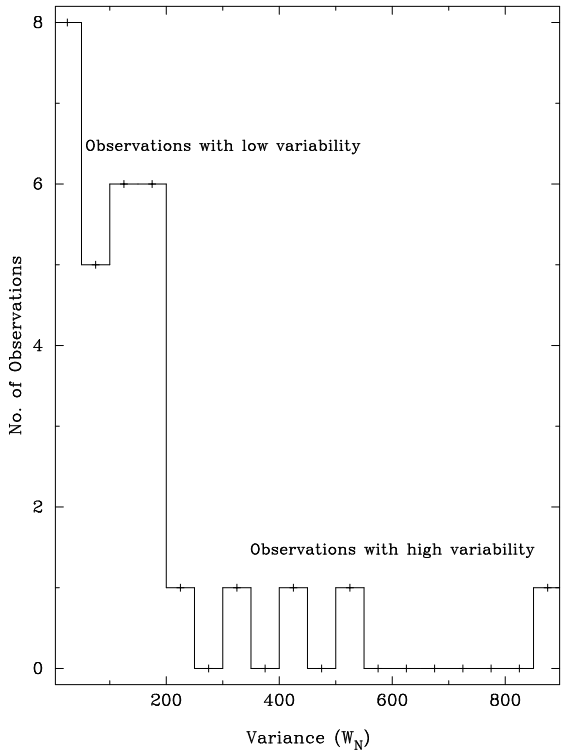}
\caption{The histogram of number of observations with respect to  the overall
variability of the
2-7 keV flux during the particular pointed observation. The x-axis gives the 
variance per bin 
(see text for details), the value of 200 in this axis
gives a limit of the low-variability range while that $>$250 corresponds to high
variability range. The lagged anti-correlation is present in all the observations
with high variability, and is also marginally observed in the observation falling in
the 200-250 bin (the \textit{marginal} case). Of all the observations with low
variability only one observation shows some hint of lagged anti-correlation (the
\textit{anomalous} case). (See Table \ref{tab1}).\label{fig1}}
\end{figure}

Here we report the presence of such an anti-correlated time lag of the hard X-ray (20
-- 50 keV) emission to that of the soft X-ray (2 -- 7 keV) emission observed in six
pointed observations by {\it RXTE-}PCA during the $\chi$ state of \grs. During these
particular pointings the `in situ' pivoting feature in the spectra is seen similar to
what reported for \x~ \citep{choudhury04apjl}. More importantly, we also report a
corresponding change in the Quasi Periodic Oscillations (QPO) parameters ubiquitously
present in the low hard state of this source, most notably being the shift in the centroid frequency.

\section{Data and Analysis}
The pointed observations of the narrow field of view instrument PCA aboard the {\it
RXTE} are used for the timing as well as the spectral analysis. The lightcurves for
the cross-correlation tests as well as the spectra (obtained from the PCA) use the
standard 2 form of data (all PCUs added), with all the procedures of data filtering,
background and deadtime corrections strictly adhered to. The lightcurve for the QPO
analysis was obtained from the single bit mode data, covering the 2--7 keV band. The
data reduction and analysis was carried out using \texttt{HEASOFT} (v5.2), which
consists of (chiefly) \texttt{FTOOLS} (V5.2), \texttt{XRONOS} (v5.19) and
\texttt{XSPEC} (V11.2).

\subsection{Data Selection}
This source is extremely variable in nature and it has been classified into several
variability classes \citep{belloni00aa}. Based on the RXTE pointed observations
carried out in 1996-1997, \citet{belloni00aa} have identified long duration steady
hard states (the `C' state in their nomenclature) as the $\chi$ class. As discussed
by \citet{mcclintock04} (see also \citealt{rao00apj} and \citealt{vadawale01aa}),
this is the closest analogue to the canonical low-hard states of Galactic black hole
sources, and it is in this state that the source seems to spend the most of its time.
This $\chi$ state is further subdivided into 4 subclasses, $\chi1$ to $\chi4$. The
subclasses $\chi1$ to $\chi3$ are the long uninterrupted steady states whereas the
$\chi4$ class is a collection of isolated individual pointed observations with
properties similar to $\chi1$ to $\chi3$ classes. Since a portion of other
variability classes (like the $\alpha$ class) can look like the $\chi$ class, we have
only taken the $\chi1$ to $\chi3$ classes for the present study. \citet{belloni00aa}
classification covers a total of 49 pointed observations from {\it RXTE-}PCA covering
the early years of its operation with the source in the $\chi$ state, of these
the $\chi$ (1--3) states consist of 30 observations (with some long observations
separated into more than one observation IDs., in the \textit{RXTE} data archives).
All these observations pertaining to the $\chi$ (1-3) state, as per the
classification covered by  \citet{belloni00aa}, was inspected for the presence of the
anti-correlated lag between the hard (20--50 keV) and soft (2--7 keV) emission,
following the analogous results obtained for Cygnus X-3 \citep{choudhury04apjl}.
Considering the fact that the mechanism causing the change in the soft X-ray spectrum
(the accretion rate being one of the the most likely candidate), withthe
corresponding delayed opposite change in the hard X-ray emission, is very unlikely to
be a continuous process, a large number of long pointed observations are needed to
serendipitously capture the occasional individual events that may have occurred
during the pointings. The 30 observations (pertaining to the $\chi$ (1--3) states)
are classified according to the variability exhibited by the source during the each
particular pointed observation. To check for the variability we obtain  the variance
from a constant fit to the lightcurve (binned to $128 s$), normalized to the number
of time bins (W$_N$). The histogram giving the number of observations as a function
W$_N$ is plotted in Figure \ref{fig1}, where it is clearly seen that for values of
W$_N$ $>$ 250 there are four observations  (occurring on MJD 50441-42, 50477, 50480,
50729) corresponding  to the high variability cases, within the class of observations
considered by us. It is interesting to note that the lagged anti-correlation of the
hard X-rays (20-50 keV) with respect to the soft X-ray (2-7 keV) is seen in all these
four  cases. In Figure \ref{fig1} only one observation corresponds to the variability
measure (W$_N$) in the range 200-250, and this band we use to demarcate the boundary
of high and low variability, and this observation gives a marginal detection of the
anti-correlated delay. Of the 25 observations corresponding to the low variability
cases (Figure \ref{fig1}), only in one case the lagged anti-correlation is seen, in
no other case there is any detection of the delay in the correlation results. The
presumption behind the classification with respect to the variability was that the
observations during which significant variability is seen will be the ones where
noticeable features of the lagged anti-correlation can be obtained. Of course this
precludes the fact that the variability should be non-monotonic in nature because,
firstly the continuous monotonic evolution will result in state transition, and
secondly, more importantly from the observational purview, a non-monotonic evolution
with a point of maxima/minima is the best feature to measure the delay in the
anti-correlation.

\begin{figure}
\figurenum{2}
\epsscale{1.20}
\plotone{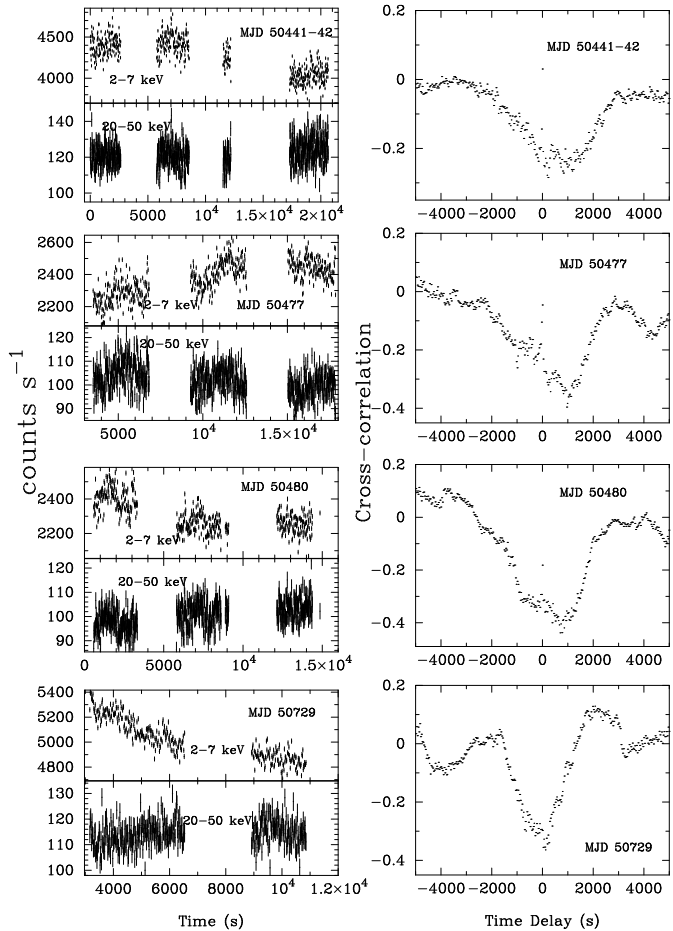}
\caption{The cross-correlation between the soft (2-7 keV) and hard (20-50 keV) X-ray
flux in the $\chi2$ (MJD 50441-42, 50477, 50480) and $\chi$3 (MJD 50729) states for
which the lagged anti-correlation is observed. These 4 observations
correspond to the high variability cases of Figure \ref{fig1}. \label{fig2}}
\end{figure}

\begin{deluxetable*}{llclcc}
\tabletypesize{\scriptsize}
\tablecolumns{6}
\tablewidth{0pc}
\tablecaption{The details of the pointed observations for which the lagged 
anti-correlation of the hard X-rays (20--50 keV, PCA) with respect to the 
soft X-rays (2--7 keV, PCA). (Here X corresponds to 20402-01).\label{tab1}}
\tablehead{
\colhead{MJD} & \colhead{Obs. Id.} & \colhead{Delay (sec) (error)} &
\multicolumn{3}{c}{ statistical correlation coefficient}\\
\cline{4-6}
\colhead{}&\colhead{}&\colhead{}&\colhead{F.F.T. (error)}&\colhead{Pearson (Null.
Prob.)}&\colhead{Spearman (Null. Prob.)}
}
\startdata
50441-42 & X-08-00, -08-01 & $\sim$220--1000$(\pm87)$ & $\sim -0.38 (\pm10^{-5})$ &
$-0.37 (\sim10^{-6})$ & $-0.38 \left( \sim10^{-6}\right) $\\
50477 & X-13-00 & $\sim960(\pm50)$ & $\sim -0.40 (\pm10^{-5})$ & $-0.48
(\sim10^{-7})$ & $-0.52 (\sim10^{-14})$ \\
50480 & X-14-00 & $\sim 704(\pm35)$ & $\sim -0.44 (\pm10^{-5})$ & $-0.48
(\sim10^{-9})$ & $-0.54 (\sim10^{-14}$) \\
50729 & X-49-00 & $\sim128(\pm30)$ & $\sim -0.36 (\pm10^{-5})$ & $-0.32
(\sim10^{-3})$ & $-0.37 (\sim10^{-9})$ \\
\hline
50436\tablenotemark{1} & X-07-00 & $\sim1600(\pm200)$ & $\sim-0.28 (\pm10^{-5})$ &
$-0.27 (\sim10^{-3})$ & $-0.24 (\sim10^{-4})$ \\
\hline
50455\tablenotemark{2} & X-10-00 & $\sim<1050$ & $\sim -0.35 (\pm0.10)$ & $-0.13
(\sim0.06)$ & $-0.18 (\sim10^{-3})$ 
\enddata
\tablenotetext{1}{marginal case}
\tablenotetext{2}{anomalous case}
\end{deluxetable*}

\subsection{Hard X-ray delay and pivoting in the spectrum} Of the observations listed
by \citet{belloni00aa} belonging to the $\chi$ (1--3) states, on four occasions
(Figure \ref{fig2}), corresponding to the high variability cases (Figure \ref{fig1}),
the cross-correlation results show the hard X-rays to be anti-correlated and lagging
with respect to the soft X-ray emission (the details are given in Table \ref{tab1}).
The cross-correlation is initially obtained using the \texttt{XRONOS} program
`crosscor', which uses the F.F.T. algorithm to compute the coefficient which is
normalized by the square root of the product of number of good newbins of the
concerned lightcurves, effectively this coefficient is the cross covariance of the
two lightcurves. On MJD 50441-42, 50477 \& 50480 the source was in the $\chi$2 state
\citep[also known as the radio-quiet  $\chi$ state -][]{vadawale01aa}, and on MJD
50729 the source was in the $\chi$3 state (the radio-loud  $\chi$ state). We further
substantiate the correlation results by performing the Pearson's test as well as
the Spearman's rank correlation test after correcting for the delay of the hard
X-ray flux for each corresponding observation, the coefficients thus obtained are
given in Table \ref{tab1}. In obtaining these coefficients, we have used data
corresponding to only those time stamps for which both the fluxes are present after
shifting the hard X-ray lightcurve to correct for the delay. In these cases if the
hard X-ray lightcurve is not compensated for the delay, the correlation coefficients
obtained are as follows; MJD 50441-42 -- Pearson's = 0.09 \& Spearman's = 0.12, MJD
50477 -- Pearson's = -0.03 \& Spearman's = 0.02, MJD 50480 -- Pearson's = -0.04 \&
Spearman's = -0.06, MJD 50729 -- Pearson's = -0.28 \& Spearman's = -0.26. Comparing
these values with the coefficients obtained for the delay compensated hard X-ray
lightcurves (table \ref{tab1}) establishes that these 4 observations pertaining to
the high variability cases, within the $\chi$ (1--3) class, do indeed exhibit this
physical phenomenon of the hard X-ray emission being delayed as well as
anti-correlated to the soft X-ray emission process. The relative closeness of the
Pearson's and Spearman's rank (anti-)correlation values in the two cases (with and
without the compensation for the delay in the hard X-rays) on MJD 50729 is easily
attributed to the fact that the lag measured in this occasion is only $128 s$.

\begin{figure}
\figurenum{3}
\epsscale{1.10}
\plotone{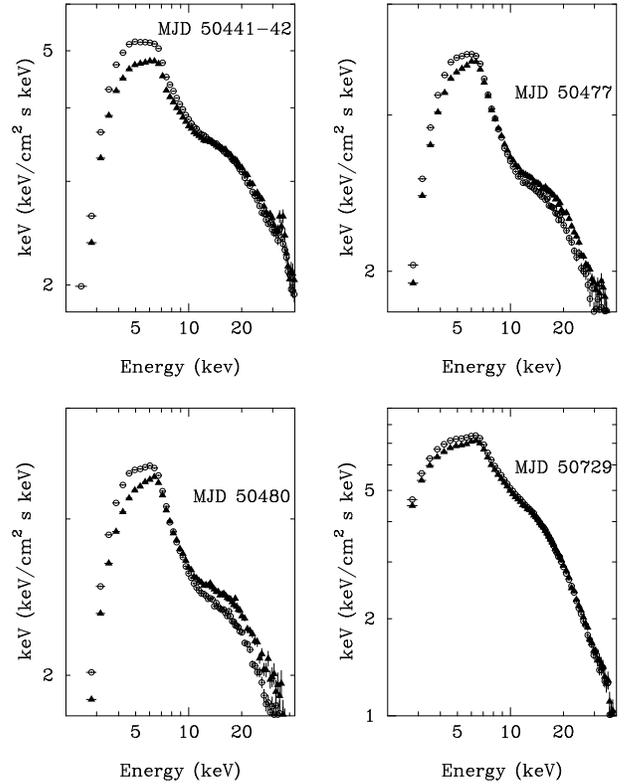}
\caption{The broadband X-ray spectra of the source on the days for which the lagged
anti-correlation is observed (Figure \ref{fig2}, for the soft and the hard regions of
the lightcurve, resulting in comparative softer and harder spectral distribution,
displaying the `in situ' pivoting in the $\chi$2 and $\chi$3 state.\label{fig3}}
\end{figure}

\begin{figure}
\figurenum{4}
\epsscale{1.10}
\plotone{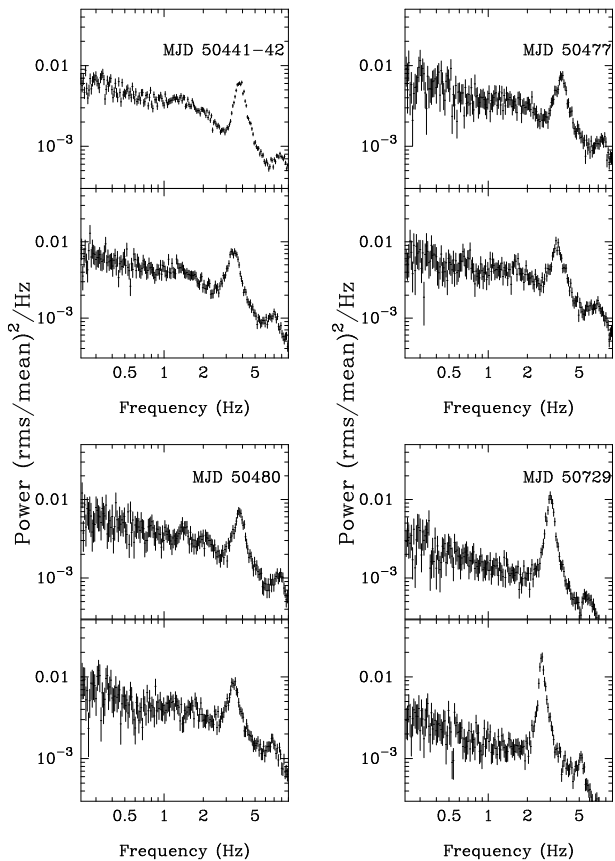}
\caption{The power density spectra of the source on the days for which the lagged
anti-correlation is observed, for the soft (top panels) and the hard (bottom panels)
regions of the lightcurve, depicting the shift of the QPO frequencies corresponding
to the soft and hard regions. \label{fig4}}
\end{figure}
 
On all these 4 days, the X-ray spectral evolution displays a pivoting pattern, during
the individual pointings. The time scale of the delay in the cross-correlation gives
the time scale for the spectra to harden after the softer flux decreases and/or
vice-versa. The interesting feature in the spectral evolution is the difference in
the pivot point on the different days, which further underlines the unique diverse
variability features of this source. In the source, the pivoting occurs around
$\sim$7 keV on MJD 50477 \& 57480 and in the range $\sim$10--20 keV on MJD 50441-42
\& 50729 (Figure \ref{fig3}, in contrast to the energy region of $\sim 20-30$ keV for
observation taken on different days \citep{choudhury03apj}. \citet{trudyulobov99astl}
\& \citet{trudyulobov01apj} have covered a large set of observations of \grs~ during
this era, which overlaps with the  $\chi$ states reported here. As reported by them,
the spectra do show pivoting at similar energy, but for observations on different
days. In Table \ref{tab1}, we report the time scale of this spectral evolution from
the cross-correlation results, from observations with in the span of one pointed
observations.
\paragraph{X-ray spectral fitting.}
We have used the model prescribed by \citet{vadawale03apj} to unfold the count
spectrum. The value of the parameters and the error bars are similar to the other
$\chi$ state spectra mentioned in \citet{vadawale03apj}. Detailed spectral analysis
is out of scope of this paper and not needed as the given results are model
independent, for the purpose of the spectral fit was to mimick the shape of the count
spectrum in order to unfold it and not to provide any physical interpretation from
the fit parameters.

\subsection{Quasi Periodic Oscillation}
Quasi periodic oscillation is an ubiquitous feature of this source in the low state.
To characterize the nature of the physical phenomenon giving rise to the lagged
anti-correlation, we have done a detailed study of the QPO characteristics,
corresponding to the comparative softer and harder regions of the lightcurve. The
power density spectra (PDS) was obtained from the single bit mode of data in the 2-7
 keV band, with a bin size of 10 ms, using the \texttt{XRONOS} program `powspec'.
This program where the PDS was computed by an FFT algorithm and the errors were
obtained by propagating the theoretical error bars of the spectra from individual
intervals (from the relevant chi-square distribution), while the normalization was
such that the integral of the spectra gives the squared rms fractional variability
(with the white noise subtracted). The QPO parameters were obtained by fitting a
powerlaw to model the continuum of the PDS and a Lorentzian to fit the QPO profile
(Figure \ref{fig4}), the details of which are given in Table \ref{tab2}. The centroid
frequency of the QPO evolves consistently with the minute change in the X-ray
emissions during the individual pointed observations. The frequency always has a
higher value at the softer region of the X-ray emission. Further, the rms variability
of the QPO is always more in the harder region of the lightcurve. This small but
indisputable and consistent change in the QPO parameters  provide further validation
of the small but definitely perceptible change in the accretion state in the system
during these episodes of soft X-ray emission impelling the hard X-ray emission, in
the steady low states. \citet{trudyulobov99astl} \& \citet{trudyulobov01apj} have
obtained the centroid frequencies of the QPO during the four $\chi$ states reported
here, averaged over the entire individual observations, the values of which indeed
lie between the two values obtained by us corresponding to the softer and harder
regions of the lightcurve (except for observation on MJD 50477, when the averaged
value coincides with the centroid frequency in the softest region of the lightcurve
-- Table \ref{tab1}).

\begin{deluxetable*}{lccccccccc}
\tablecolumns{10}
\tablewidth{0pc}
\tablecaption{The details of the quasi-periodic oscillations (QPO) in the soft and
hard regions of the lightcurves.\label{tab2}}
\tablehead{
\colhead{} & \colhead{} & \multicolumn{2}{c}{Centroid frequency, $\nu$(Hz)} &
\colhead{} &  \multicolumn{2}{c}{FWHM} & \colhead{} & \multicolumn{2}{c}{RMS
fluctuation}\\
\cline{3-4} \cline{6-7} \cline{9-10}
\colhead{MJD} & \colhead{} & \colhead{soft} & \colhead{hard} & \colhead{} &
\colhead{soft} & \colhead{hard} & \colhead{} & \colhead{soft} & \colhead{hard}
}
\startdata
50441-42 & \vline & $3.85 (\pm0.01)$ & $3.43 (\pm0.02)$ & \vline & 0.77 & 0.83 &
\vline & 8.18 & 9.3 \\
50477 & \vline & $3.63 (\pm0.03)$ & $3.39 (\pm0.03)$ & \vline & 0.64 & 0.58 & \vline
& 7.36 & 7.61 \\
50480 & \vline & $3.79 (\pm0.03)$ & $3.41 (\pm0.02)$ & \vline & 0.5 & 0.48 & \vline &
6.82 & 7.75 \\
50729 & \vline & $3.02 (\pm0.01)$ & $2.60 (\pm0.01)$ & \vline & 0.36 & 0.26 & \vline
& 7.60 & 7.80 \\
\hline
50436\tablenotemark{1} & \vline & $3.18 (\pm0.02)$ & $3.07 (\pm0.01)$ & \vline & 0.68
& 0.70 & \vline & 9.16 & 9.64 \\
\hline
50455\tablenotemark{2} & \vline & $2.98 (\pm0.02)$ & $2.76 (\pm0.02)$ & \vline & 0.38
& 0.53 & \vline & 8.23 & 8.74 \\
\enddata
\tablenotetext{1}{marginal case}
\tablenotetext{2}{anomalous case}
\end{deluxetable*}

\subsection{Marginal and anomalous cases}
\begin{figure}
\figurenum{5}
\epsscale{1.20}
\plotone{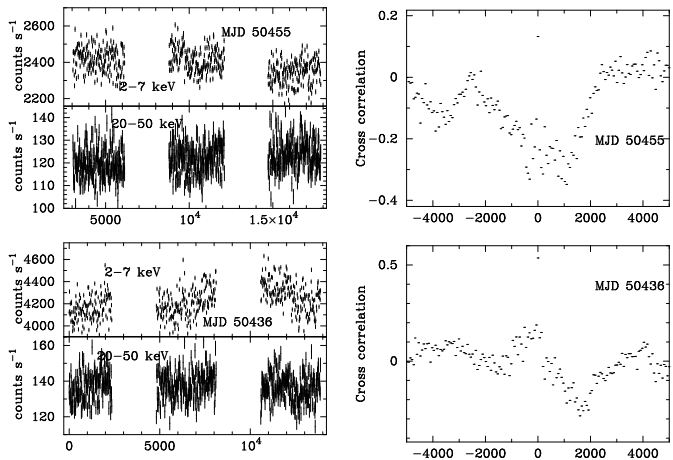}
\caption{The cross-correlation between the soft (2-7 keV) and hard (20-50 keV) X-ray 
flux for the \textit{marginal} (MJD 50436, top) and \textit{anomalous} (MJD 50455,
bottom) cases.\label{fig5}}
\end{figure}

\begin{figure}
\figurenum{6}
\epsscale{1.10}
\plotone{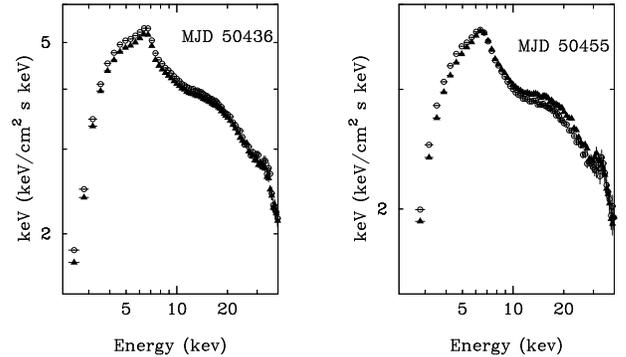}
\caption{The broadband X-ray spectra of the source on the days for the
\textit{marginal} (MJD 50436) and \textit{anomalous} (MJD 50455) cases for which the
lagged anti-correlation is observed, for the soft and the hard regions of the
lightcurve, resulting in comparative softer and harder spectral distribution,
displaying the `in situ' pivoting in the $\chi$2 state.\label{fig6}}
\end{figure}

\begin{figure}
\figurenum{7}
\epsscale{1.10}
\plotone{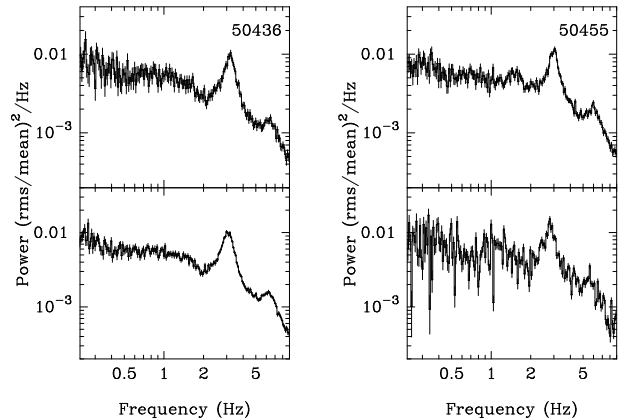}
\caption{The power density spectra of the source on the days for the
\textit{marginal} (MJD 50436) and \textit{anomalous} (MJD 50455) cases for which the
lagged anti-correlation is observed, for the soft (top panels) and the hard (bottom
panels) regions of the lightcurve, depicting the shift of the QPO frequencies
corresponding to the soft and hard regions. \label{fig7}}
\end{figure}

In Figure \ref{fig1} the low variability and high variability cases are demarcated by
the histogram bin corresponding to W$_N$ value of 200-250 which has only one
observation (MJD 50436), and indeed there is a  detection of lagged anti-correlation
with a comparative smaller value of coefficient, and we classify this as the
\textit{marginal} case of detection of lagged anti-correlation. The lightcurve and
the cross-correlation results are shown in Figure \ref{fig5} (bottom panels). As is
the normal case with the other observations without the lagged anti-correlation, at
zero delay it shows a fairly strong positive correlation (but not so strong as the
normal lower variability cases) but it shows a definite, albeit comparatively
weaker, anti-correlation at a delay of $1600 s$. The spectra shows a weak pivoting
at $\sim20$ keV (Figure \ref{fig6}), and also
the QPO parameters vary according to the so far established pattern (Figure
\ref{fig7}), with the centroid frequency shifting by 0.1 Hz, whereas for the
other high variability cases the shift is of the order of $\sim0.5$ Hz (Table
\ref{tab2}). This particular observation does indicate a gradual progression of the
physical properties and phenomena which may be classified by the empirical measure of
variability quantified in Figure \ref{fig1}.

Of the 25 cases of the low variability observations, only in one case the presence of
the lagged anti-correlation is seen (on MJD 50455). Since this marks as an exception
to the general behaviour, we classify this particular observation as
\textit{anomalous} case. The cross-correlation, depicted in Figure \ref{fig5} (top
panels), shows anti-correlation of the hard X-rays (20-50 keV) with respect to soft
X-rays (2-7 keV), with a delay that is fairly widely spread, albeit with as strong
and significant cross correlation coefficient as the high variability cases, but the
Pearson and Spearman's rank coefficient (after compensating the for a delay of $1000
s$ in the hard X-ray emission) are significantly weaker.

\section{Discussion}
In \citet{choudhury03apj} it was shown that the characteristics of X-ray emission in
the hard state of the Galactic black hole binary systems seem to possess universal
features, leading to the picture of a truncated accretion disk system. Diverse
sources, viz. \x, \grs, Cygnus X-1 \& GX 339-4 showed similar pivoting in X-ray
spectrum, with the pivot point varying for the different sources. This pivot point
was considered to be dependent on the extent of the truncation radius. In this
geometrical structure, the Compton scattered X-rays originate from a region close to
the compact object inside the truncated disk, giving rise to the non-thermal
component in the hard X-ray regime. The extent of the accretion disk determines the
amount of seed photons available for the Comptonizing component. With any change in
the extent of the truncation radius (probably due to change in accretion rate) there
is an opposite change in the flux of the Comptonizing component in the spectrum,
leading to the anti-correlation between the hard and soft X-rays. This physical
picture is applicable for all the models that require a truncated disk, viz. ADAF
\citep{narayan94apjl} or TCAF \citep{chakrabarti96phr}. 

\citet{choudhury04apjl} discovered a lagged anti-correlation between the hard and
soft X-rays in Cygnus X-3, roughly the origins of which can be ordained to be from
the inner Comptonizing cloud and the thermal disk, respectively. This delay may be
attributed to the viscous time scale of matter in the truncated thermal disc, which
is the readjustment time scale of the disk as well as the inner Comptonizing cloud.
In this scenario, the soft X-ray flux changes due to any inherent accretion process
(viz. the accretion rate), and after a lag of the observed time scale the disk
readjusts its geometrical size effecting an opposite change in the Comptonizing cloud
in the inner regions, and thence the non-thermal component of the spectra changes,
resulting in the pivoting phenomenon.

Due to the similarity of the features of the X-ray emissions of \grs~ with that of
\x~ in the hard state we were prompted to search for analogous lagged
anti-correlation of the hard X-ray emission (20-50 keV) with respect to the soft
X-ray emission (2-7 keV) resulting in the pivoting in the spectrum. Inspection of all
the observations compiled by \citet{belloni00aa} yielded 4 occasions when the pointed
{\it RXTE}-PCA observations were able to catch the serendipitous events leading to the
restructuring of the accretion system. The difference in the pivot point in the
spectra for $\chi$3 and $\chi$2 (sub)states suggest that the intrinsic accretion system
\grs~ is more complex than that of \x~ with the broad scenario of the accretion
system in the two $\chi$ (sub)states being not exactly the same. But the short time
scale evolution of the accretion system in the X-ray hard state, resulting in the
readjustment in the disk structure and size, is similar for both the (sub)states.

The presence of low-frequency QPO in \grs~ \citep{muno99apj} is an inherent feature
of the X-ray hard state \citep{rao01aaps}. Various prescriptions exist without any
general consensus as to the precise nature of this temporal characteristic. These
low-frequency QPOs are found to be correlated to observed phase-lag in the Fourier
cross-spectrum for two different channels \citep{reig00apj} which may arise from the
Compton spectrum in the inner regions of the disk \citep{nobili00apjl}. Further, the
centroid frequency is shown to be correlated to the spectral index (of the SED) in
the hard $\chi$ state \citep{vignarca03aa}. \citet{rao00aa} found the low-frequency
QPO in the $\chi$3 state to originate in the region of the Comptonizing cloud, while
\citet{chakrabarti00apjl} interpret the QPO as due to oscillation of the region
responsible for the hard X-ray emission. Arguing on a similar vein
\citet{titarchuk04apj} provide a two component model with a truncated thermal
accretion disk with a quasi-spherical `transition layer' inside (TL model) giving
rise to the hard X-ray emission as well as the QPO. The general consensus seems to
converge toward the idea that a physical model for the compact corona must naturally
produce oscillations on time scales of hours to days \citep{remillard04aip}. Models
not conforming directly to this geometrical structures, viz. magnetic flood scenario
\citep{tagger99aa} leading to `accretion ejection instability' \citep{tagger04apj}
also invoke advection as a necessary element of the theoretical formalism, and the
QPOs are supposed to originate at inner regions of the accretion. The consistent
increase in the centroid frequency of the QPO with the softening of the X-ray
emission, in the small scale reported here, provides additional support as well as
constraint on the physical scenario favoring a truncated disk and high energy inflow
inside. This increase in frequency may indicate the decrease in truncation radius,
resulting in a physically and geometrically bigger accretion disc (rendering the
increased softness of the spectrum). Also, the decrease in the rms variability of the
QPO with increasing softness may indicate the decrease in volume of the Comptonizing
cloud, under the assumption that the physical origin of the QPO phenomenon lies in
this highly energetic matter. 

The discovery of the hard X-ray delay (anti-correlated) in the hard state of this
source (second after \x) represses the accretion models without a truncated disk. In
addition to the pivoting in the spectra, the consistent increase in the QPO centroid
frequency puts the observational feature on very firm footing. Further search for
such features in sources like Cygnus X-1 and GX 339-4, where the pivot point is at higher energy, 50--80 keV \& $>$300 keV, respectively, is needed to get a complete
picture of the X-ray emission in the hard state.

\section*{Acknowledgments} This research has made use of data obtained through
the HEASARC Online Service, provided by the NASA/GSFC, in support of NASA High
Energy Astrophysics Programs.

\end{document}